\documentstyle{article}
\begin{document}

\begin{center}
{\LARGE \bf Coherent brightness modulations in the \\ dwarf nova AT~Cancri}
\vspace{0.5cm}

{\Large \bf Albert Bruch}
\vspace{0.1cm}

Laborat\'orio Nacional de Astrof\'{i}sica, Rua Estados Unidos, 154, \\
CEP 37504-364, Itajub\'a - MG, Brazil
\vspace{0.3cm}

{\Large \bf James Boardman}
\vspace{0.1cm}

CBA Wisconsin, Luckydog Observatory, 65027 Howath Road, de Soto, WI 54624, USA
\vspace{0.3cm}

{\Large \bf Lewis M. Cook}
\vspace{0.1cm}

American Association of Variable Star Observers, 1739 Helix Ct. 
Concord, CA 94518, USA
\vspace{0.3cm}

{\Large \bf Michael J.\ Cook}
\vspace{0.1cm}

CBA Ontario, Newcastle Observatory, 9 Laking Drive,
                 Newcastle, ON L1B 1M5, Canada
\vspace{0.3cm}

{\Large \bf Shawn Dvorak}
\vspace{0.1cm}

AAVSO, Rolling Hills Observatory, 1643 Nightfall Drive,
                 Clermont, FL 34711, USA
\vspace{0.3cm}

{\Large \bf James L.\ Jones}
\vspace{0.1cm}

CBA Oregon, Jack Jones Observatory, 22665 Bents Road NE,
                 Aurora, OR 97002, USA
\vspace{0.3cm}

{\Large \bf John W.\ Rock}
\vspace{0.1cm}

CBA Wilts, 2 Spa Close, Highworth, Swingdon, SN6 7PJ, UK
\vspace{0.3cm}

{\Large \bf Geoffrey Stone}
\vspace{0.1cm}

CBA Sierras, Sierra Remote Observatories, 
                 44325 Alder Heights Road, Auberry, CA 93602, USA
\vspace{0.3cm}

{\Large \bf Joseph H.\ Ulowetz}
\vspace{0.1cm}

CBA Illinois, Northbrook Meadow Observatory, 855 Fair Lane,
                 Northbrook, IL 60062, USA
\vspace{0.5cm}

(Published in: New Astronomy, Vol.\ 66, p.\ 22 -- 28 (2019))

\vspace{0.5cm}
\end{center}

\begin{abstract}
Light curves of the Z~Cam type dwarf nova AT~Cnc observed during standstill in
2016 and 2018 are analyzed. On the time scale of hours, previous reports
on periodicities, in particular the presence of negative superhumps, could
not be confirmed. Instead, a modulation with a period equal to the 
spectroscopic orbital period was detected which we thus interpret as a
manifestations of the binary revolution. It enables us to derive a more 
accurate value of 0.201634 $\pm$ 0.000005 days (or its alias of 0.021580 
days) for the period. AT~Cnc also exhibits a hitherto unreported modulation 
of 25.731 $\pm$ 0.005 min, stable in period but not in amplitude over the 
entire time base of two years of the observations. We tentatively interpret 
this modulation in the context of an intermediate polar model for the system.
\vspace{1ex}

{\parindent0em Keywords:
Stars: binaries: close --
Stars: novae, cataclysmic variables --
Stars: individual: AT~Cnc}
\end{abstract}

\section{Introduction}
\label{Introduction}

Cataclysmic variables (CVs) are among the most variable objects in the sky. 
They exhibit variations of their brightness at practically all wavelengths
and on time scales ranging from seconds to millennia. Almost all of these are 
directly or indirectly caused by mass transfer from a low mass late 
type star (the secondary) to a more massive white dwarf (the primary) in 
close orbit around each other. For an exhaustive introduction to our 
understanding of CVs see, e.g., Warner (1995).

In the majority of these systems, unless the white dwarf has a strong
magnetic field which guides mass directly to its surface, the transferred 
matter first forms an accretion disk around the more massive star where
it loses angular momentum and can then settle down on the central
object. It is thought that depending on the detailed conditions this disk 
can switch between a low viscosity and low brightness state termed 
quiescence with little mass accretion on the white dwarf, and a high 
viscosity state which permits dumping much of its matter onto the star 
in a short time. A detailed review of this so-called disk-instability
model is given by Lasota (2001). The release of gravitational energy 
during this process then causes a brightening 
of the system of several magnitudes (i.e., an outburst) which can last 
from a few days to a couple of weeks. These systems are known as dwarf 
novae. A small subgroup of them can remain occasionally in an intermediate 
state between a full-fledged outburst and quiecence. Such states are termed 
standstills. After their prototype, the CVs belonging to this subgroup are 
called Z~Cam stars (Buat-M\'enard et al. 2001, Simonsen et al. 2014).

One of these is AT~Cnc, originally detected as a variable
star by Romano \& Perssinotto (1968). The early history of the system is
summarized by Nogami et al.\ (1999) to whom the reader is referred for
details. The long term light curve is quite well documented in a
series of short communications in particular by Meinunger (1981)
and Goetz (1983, 1985, 1986, 1988a, 1988b, 1990, 1991). 
From the 1980s on AT~Cnc has also been 
extensively monitored by amateur astronomer as is testified by its
decades long light curve stored in the American Association of Variable
Star Observers (AAVSO) International Database\footnote{https://www.aavso.org}. 
Like many other
Z~Cam stars, AT~Cnc is a very active system which remains hardly ever 
in quiescence but exhibits a rapid succession of outbursts, sometime
interrupted by standstills. Shara et al.\ (2012) detected a shell around 
the system, i.e., the remnant of a nova eruption which occurred three 
or four centuries ago (Shara et al. 2017).

While the long term behaviour of the star is thus well known, more detailed
studies of AT~Cnc are rare. Only Nogami et al.\ (1999) published time
resolved spectroscopic
observations, obtained during a standstill. They measured an orbital period 
of $P_{\rm orb} = 0.2011 \pm 0.0006$ d ($4.83 \pm 0.01$ h). They also tried to
determine the primary star mass ($M_1 = 0.9 \pm 0.5 M_\odot$), but the large
error causes it to be hardly of any use. The orbital inclination is
definitely low. Depending on the origin of the H$\alpha$ emission line
Nogami et al.\ (1999) obtained either $17^{\rm o} \pm 3^{\rm o}$, if the
emission comes from the secondary star, or $36^{\rm o} \pm 12^{\rm o}$ if
it arises in the accretion disk. The spectrum also exhibits the Na~I~D
absorption lines. This would normally be attributed to an imprint of the
secondary star. If AT~Cnc would have been in quiescence this would not be 
surprising considering the rather long orbital period. But this interpretation
is less obvious during a standstill. The radial velocity variations of
the emission and aborption lines are in phase. If the Na~I lines really 
come from the secondary, this would strengthen the case of the H$\alpha$
emission to arise from the same source. But the width of H$\alpha$ emission 
lines from the secondary, when seen in CVs, is in general quite small, in 
contrast to what is seen in AT~Cnc (see Fig.~7 of Nogami et al. 1999). 
Moreover, the H$\alpha$ emission has a P~Cyg profile, indicating the 
presence of absorbing material in the system which may also cause the 
Na~I absorption. The Na~D lines in the spectrum of AT~Cnc are thus not 
necessarily caused by the secondary star. In a similar case, 
Ratering et al.\ (1993) observed Na~D absorption in outburst spectra of another
Z~Cam star, KT~Per, during outburst, but not in quiescence. A further
example is the SW~Sex star PX~And (Thorstensen et al. 1991). Finally,
Smith (1997) also saw Na~D absorption in their spectrum of AT~Cnc
observed during standstill, but noted the absence of TiO bands which 
might be expected to be visible if the Na-D lines were due to the secondary 
star. Thus, the origin of the H$\alpha$ emission and the Na~D absorption 
remains open.

Nogami et al.\ (1999) also obtained a limited amount of time resolved
photometric observations. They found indications for the presence of low
amplitude variations with a possible period of 0.249 or 0.132 d (or
aliases thereof) in their light curves. This is different from a period
of 0.239 d claimed to be present in AT~Cnc by Goetz (1986) and
none is close to the orbital period. A somewhat more extensive set
of photometric observations, also obtained during standstill, was
presented by Kozhevnikov (2004). He could not confirm the
variations found by Goetz (1986) and Nogami et al.\ (1999)
but, separating his data into two sets adjacent in time, he instead 
finds indications for quasi periods of 4.65 and 4.74 h, slightly
less than the spectroscopic orbital period, with amplitudes of 5 -- 9 mmag.
He interprets these variations as negative superhumps, i.e., caused by
brightness variations due to the nodal precession of an accretion disk
which is inclined with respect to the orbital plane of the binary system.
Kozhevnikov (2004) also detected a broad hump at frequencies of
0.4 -- 0.7 mHz (corresponding to periods between $\sim$25 -- 40 min)
in the power spectra of his light curves.
 
Here, we investigate additional light curves of AT~Cnc in order to address 
the question of the presence or not of consistent variations in the system.
In Sect.~\ref{Observations} the data are presented. 
In Sect.~\ref{Orbital variations} we look for variations on the orbital 
time scale, not finding any indication for the presence of superhumps, 
but a modulation which permits to refine $P_{\rm orb}$.
In Sect.~\ref{A coherent 26.7 min variation} we show the presence of a 
coherent 26.7 min variation. Finally, our results are discussed in 
Sect.~\ref{Discussion} and briefly summarized in Sec.~\ref{Conclusions}.
 
\section{Observations}
\label{Observations}

\begin{figure}
\parbox[]{0.1cm}{\epsfxsize=14cm\epsfbox{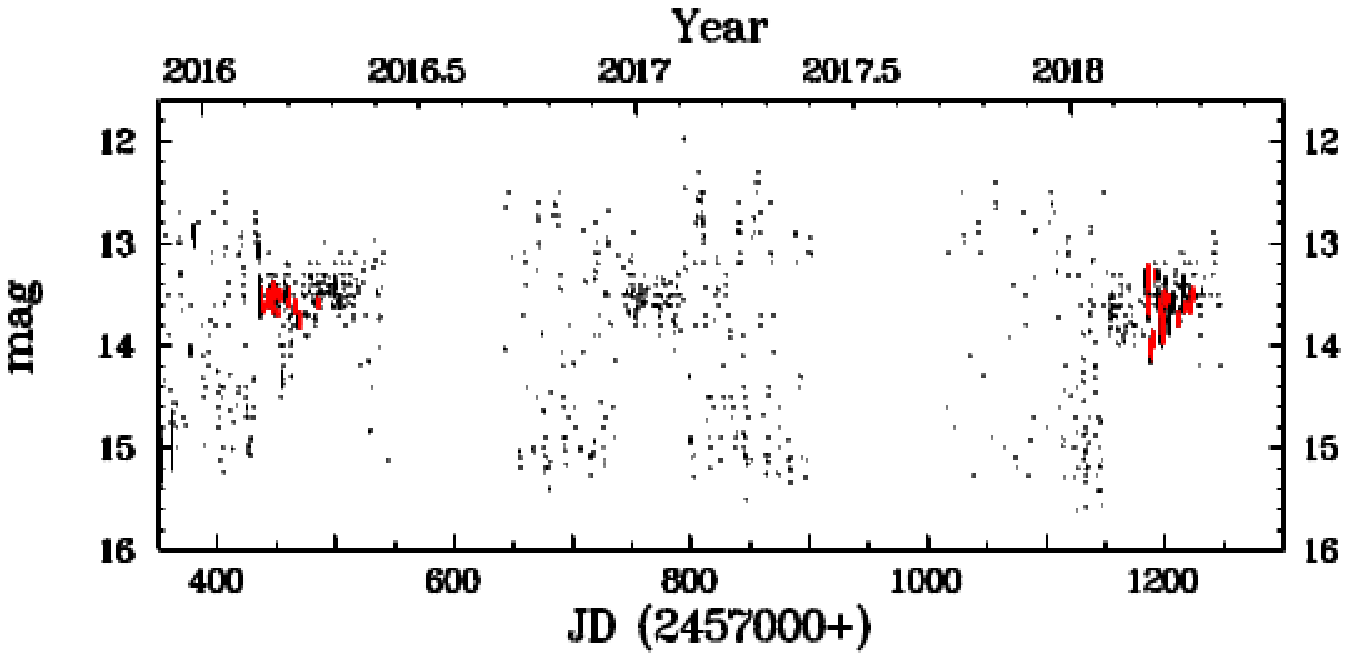}}
      \caption[]{AAVSO light curve of AT~Cnc encompassing the observing seasons
                 from 2016 to 2018. The data used in the present study
                 are shown in red. (For interpretation
                 of the references to colour in this figure legend, the
                 reader is referred to the web version of this article.)}
\label{long-term}
\end{figure}

The data used here were all obtained during
two episodes of standstill in AT~Cnc in 2016, February -- April
and 1918, March -- April. They consist of a total of 37 light curves observed 
in the Johnson $V$ band or in white light and then reduced to $V$ at a 
variety of small observatories of the Center of Backyard Astrophysics
network\footnote{https://cbastro.org}. The multiple instruments and data
reduction procedures imply that the magnitude scale may differ from one
data set to the other. However, only relative magnitude measurements are
relevant here. Thus, the scale differences are of no consequence.
Moreover, several comparison and check stars have been used to determine
the magnitude of AT~Cnc. This practically eliminates the risk to confound
variations in a comparison star which those of the target star
(Kozhevnikov 2003, 2007). A list of the employed light curves is given in 
Table~\ref{Journal of observations}. The AAVSO long term light curve
encompassing the observing seasons from 2016 -- 2018 is shown in 
Fig.~\ref{long-term}. The red dots represent the data used for the present 
study. All light curves are available at the AAVSO International 
Database.

\begin{table}

\caption{Journal of observations}
\label{Journal of observations}
\hspace{1ex}

\begin{tabular}{l@{\hspace{1ex}}*{5}{c@{\hspace{2ex}}}l}
\hline
Date & \multicolumn{2}{c@{\hspace{1ex}}}{Time (UT)} & Number of& Time resol. 
     & Oscillation$^*$ & Observer \\
     & Start      & End              & integrations & (sec) &    &     \\
\hline
2016 Feb           18 & \phantom{0}0:26 & \phantom{0}4:47 & 187 & 74 & -- &
     J.\ Ulowetz \\
2016 Feb           19 & \phantom{0}1: 4 & \phantom{0}3:52 & 144 & 71 & -- &
     S.\ Dvorak \\
2016 Feb           23 & \phantom{0}2:53 &           11:15 & 675 & 44 & +  &
     J.\ Jones \\
2016 Feb           26 & \phantom{0}2:59 & \phantom{0}6: 4 & 252 & 44 & +  &
     J.\ Jones \\
2016 Feb           27 & \phantom{0}1:21 & \phantom{0}6: 5 & 222 & 75 & +  &
     J.\ Boardman \\ [1ex]
2016 Feb           28 & \phantom{0}0:29 & \phantom{0}8:15 & 333 & 74 & +  &
     J. Ulowetz \\
2016 Mar \phantom{0}2 & \phantom{0}1:15 & \phantom{0}5:38 & 212 & 74 & +  &
     J.\ Boardman \\
2016 Mar \phantom{0}3 & \phantom{0}0: 2 & \phantom{0}6:46 & 234 & 99 & +  &
     K.\ Menzies \\
2016 Mar \phantom{0}4 & \phantom{0}5: 8 & \phantom{0}7:56 & 111 & 74 & +  &
     J.\ Ulowetz \\
2016 Mar           11 & \phantom{0}1:40 & \phantom{0}6:57 & 254 & 75 & o  &
     J.\ Boardman \\ [1ex]
2016 Mar           12 & \phantom{0}0:46 & \phantom{0}7:25 & 300 & 74 & +  &
     J.\ Ulowetz \\
2016 Mar           17 & \phantom{0}3:21 & \phantom{0}9:50 & 448 & 44 & -- &
     J.\ Jones \\
2016 Mar           21 & \phantom{0}2:44 & \phantom{0}6:50 & 183 & 74 & o  &
     J.\ Ulowetz \\
2016 Apr \phantom{0}5 & \phantom{0}4: 4 & \phantom{0}7:38 & 325 & 35 & +  &
     L.\ Cook \\
2018 Mar \phantom{0}6 & \phantom{0}3:15 & \phantom{0}9:24 & 450 & 44 & o  &
     G.\ Stone \\ [1ex]
2018 Mar \phantom{0}7 & \phantom{0}3:18 & \phantom{0}9:14 & 418 & 45 & +  &
     G.\ Stone \\
2018 Mar \phantom{0}8 & \phantom{0}3:22 & \phantom{0}9:12 & 331 & 51 & o  &
     G.\ Stone \\
2018 Mar \phantom{0}8 &           20:42 &           23:27 & 263 & 36 & -- &
     J.\ Rock \\
2018 Mar           10 &           19:47 &           22:27 & 249 & 36 & -- &
     J.\ Rock \\
2018 Mar           11 &           19:19 &           22:52 & 166 & 69 & -- & 
     D.\ Barrett \\ [1ex]
2018 Mar           12 & \phantom{0}0:30 & \phantom{0}3:28 & 148 & 73 & o  &
     M.\ Cook \\
2018 Mar           13 &           19:12 &           24: 9 & 476 & 36 & -- &
     J.\ Rock \\
2018 Mar           17 & \phantom{0}0:31 & \phantom{0}6:50 & 428 & 50 & +  &
     S.\ Dvorak \\
2018 Mar           18 & \phantom{0}0:31 & \phantom{0}6:30 & 407 & 51 & +  &
     S.\ Dvorak \\
2018 Mar           19 & \phantom{0}0:30 & \phantom{0}5:33 & 302 & 52 & +  &
     S.\ Dvorak \\ [1ex]
2018 Mar           19 &           19:14 &           22: 2 & 176 & 37 & +  &
     J.\ Rock \\ 
2018 Mar           20 & \phantom{0}0: 5 & \phantom{0}4:56 & 238 & 73 & +  &
     M.\ Cook \\
2018 Mar           21 & \phantom{0}0:47 & \phantom{0}6:35 & 395 & 51 & -- &
     S.\ Dvorak \\
2018 Mar           21 &           19:24 &           21:28 & 200 & 36 & -- &
     J.\ Rock \\
2018 Mar           23 & \phantom{0}1:46 & \phantom{0}5:27 & 326 & 40 & +  &
     S.\ Dvorak, J.\ Boardman \\ [1ex]
2018 Mar           24 & \phantom{0}0:36 & \phantom{0}3:35 & 172 & 50 & -- &
     S.\ Dvorak \\
2018 Apr \phantom{0}2 & \phantom{0}1:44 & \phantom{0}6:34 & 381 & 44 & o  &
     J.\ Boardman \\
2018 Apr \phantom{0}7 & \phantom{0}1:51 & \phantom{0}6:11 & 351 & 44 & +  &
     J.\ Boardman \\
2018 Apr \phantom{0}8 & \phantom{0}1:51 & \phantom{0}6:12 & 346 & 45 & +  &
     J.\ Boardman \\
2018 Apr           12 & \phantom{0}1:21 & \phantom{0}5:10 & 255 & 50 & +  &
     S.\ Dvorak \\ [1ex]
2018 Apr           13 & \phantom{0}1:37 & \phantom{0}5: 5 & 233 & 51 & +  &
     S.\ Dvorak \\
2018 Apr           14 & \phantom{0}1:25 & \phantom{0}4:60 & 224 & 50 & +  &
     S.\ Dvorak \\
\hline                                                            
 &  &  &  &  &  & \\
  & & \multicolumn{1}{r}{$^*$:} & 
                      \multicolumn{2}{l}{+ = definitely present} & & \\
  & &               & \multicolumn{2}{l}{o = probably present}   & & \\
  & &               & \multicolumn{2}{l}{-- = absent}            & & \\
\end{tabular}

\end{table}

\input epsf

The data analysis was performed using the MIRA software system
(Bruch 1993). In particular, timing analysis of the data
employing Fourier techniques was done using the Lomb-Scargle 
algorithm (Lomb 1976, Scargle 1982, Horne \& Baliunas 1986), well
suited for non-equidistant data.
The terms ``power spectrum'' and ``Lomb-Scargle periodogram'' are
used synonymously hereafter. Whenever light curves
from different nights were combined, the barycentric correction was
applied to the time information, using the on-line tool of
Eastman et al.\ (2010), in order to remove light travel effects
in the solar system.

\section{Orbital variations}
\label{Orbital variations}

\begin{figure}
\parbox[]{0.1cm}{\epsfxsize=14cm\epsfbox{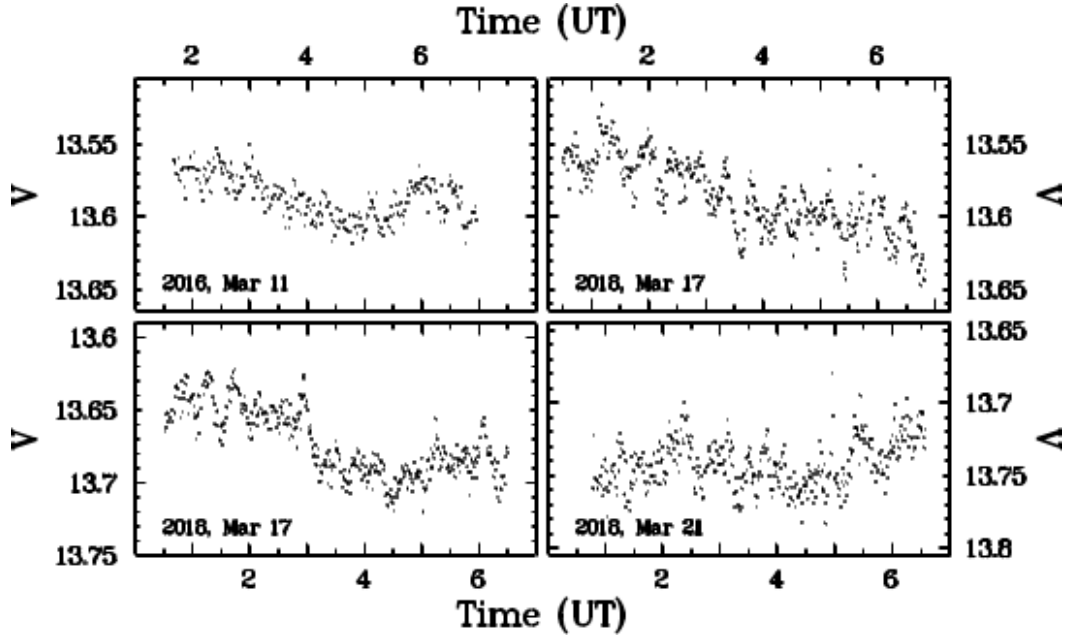}}
      \caption[]{Some representative nightly light curves of AT~Cnc,
                 all plotted on the same time and magnitude scale.}
\label{lightcurves}
\end{figure}

The light curves of AT~Cnc exhibit variations on the time
scale of hours. However, regarding them individually, no consistent
common time scale is apparent. Formal fits of simple sine functions
yield widely varying periods. In Fig.~\ref{lightcurves} we plot
representative light curves from four nights, each covering a time base
of approximately the orbital period. While significant variations on
time scales of several hours are clearly present, at most the observations
of 2016, March 11 and 2018, March 17 suggest a repeatable pattern.

In order to investigate the question of consistent variations on this
time scale further, we combined the light curves into two data sets,
one for each of the two observing seasons. Before doing so, we subtracted
the average magnitude of each light curve in order to remove night-to-night
variations and differences of the magnitude scales caused by the variety
of instruments used for the observations. A Lomb-Scargle periodogram was
then calculated from the resulting seasonal light curves.

\begin{figure}
\parbox[]{0.1cm}{\epsfxsize=14cm\epsfbox{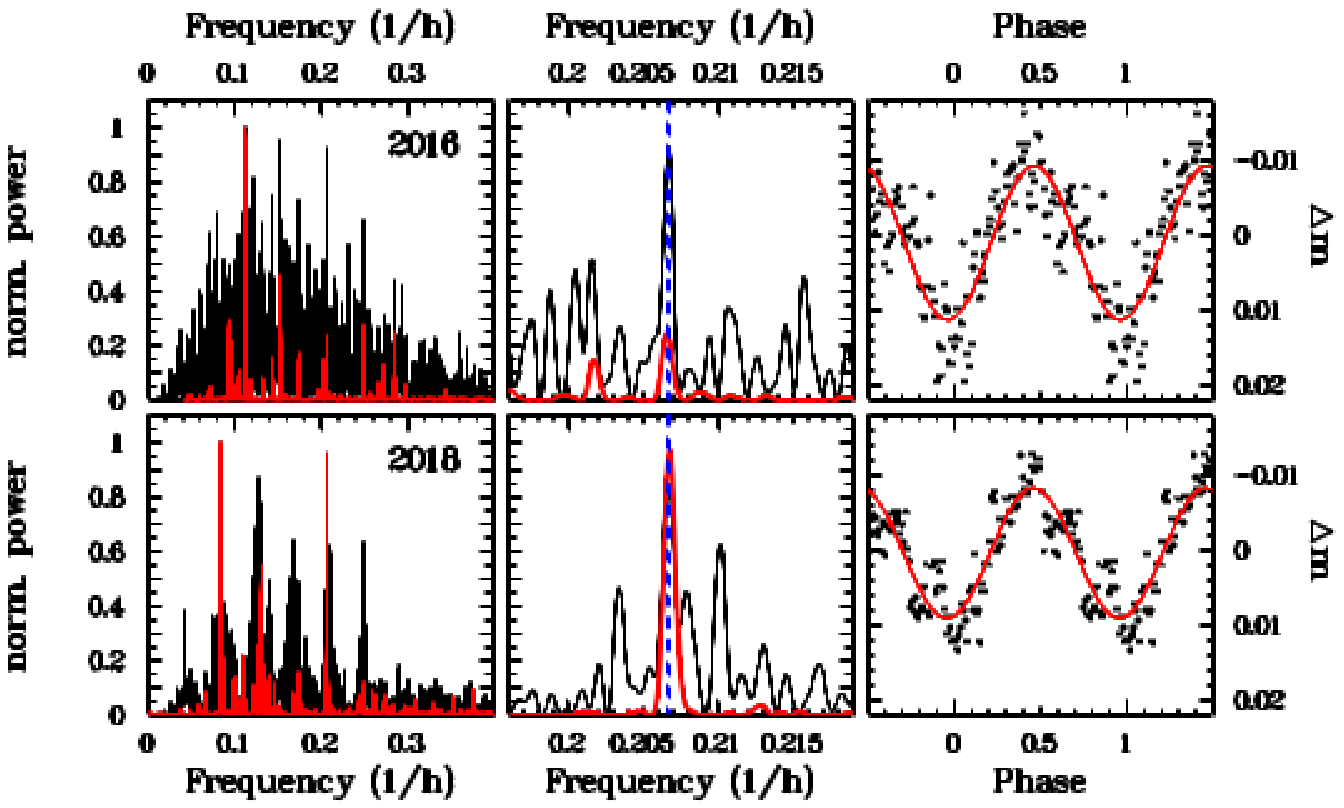}}
      \caption[]{Analysis of variations in AT~Cnc occurring on the time 
                 scale of several hours. 
                 {\it Left:} Lomb-Scargle periodograms of the
                 combined light curves of 2016 (top) and 2018 
                 (bottom), normalized to the power of the highest peak.
                 The cleaned power spectra are over plotted in red. 
                 {\it Center:} The same periodograms as shown in the left
                 frame, restricted to a narrow range around
                 the frequency corresponding to the orbital period.
                 The broken vertical line indicates the average of the
                 these frequencies measured in the two observing seasons.
                 {\it Right:} The combined light curves, folded on the
                 period of the highest peak in the central frames, binned
                 in phase bins of width 0.01. The red curves are the best
                 sine fits. (For interpretation
                 of the references to colour in this figure legend, the
                 reader is referred to the web version of this article.)}

\label{orb-period}
\end{figure}

These power spectra are plotted in the left frames 
Fig.~\ref{orb-period} as black graphs. The uneven time
distribution of the light curves gives rise to a very complicated
pattern of peaks, caused by aliasing effects and possibly the presence
of multiple periods or quasi-periods. Comparing the periods corresponding
to the most prominent peaks to those mentioned by Goetz (1986),
Nogami et al.\ (1999)
and Kozhevnikov (2004), we find only one of them, in 2018, to
be close to the 0.249 day period claimed by Nogami et al.\ (1999).
The difference between their and our period value is only 2.1 min. The
period error in our data is 0.8 min as judged from the 
standard deviation of a Gaussian fit to the respective power spectrum 
peak. Nogami et al.\ (1999) provide no error estimate, but it should even
be larger considering the short time base of their observations. Therefore,
the difference is insignificant. There may thus be some evidence that during
2018 a similar modulation as seen by Nogami et al.\ (1999) in 1997 was
present, but it cannot be considered sufficient to claim a definite
detection (see also below).

In an effort to clean the complex power spectra from alias and window
effects, we applied the CLEAN algorithm (Roberts et al. 1987) to the
data. The cleaned power spectra are superposed upon the Lomb-Scargle
periodograms as red graphs in Fig.~\ref{orb-period}. While this procedure
is not perfect (some maxima in the resulting spectra can still readily
be identified as being caused by the window function\footnote{The higest
peak in the 2018 power spectra corresponds, e.g., to a period of exactly
12~h and is thus very likely an effect of data sampling.}), the cleaned 
spectra are much easier to interpret. As an aside we first mention that
the maximum corresponding to the period claimed by Nogami et al.\ (1999)
(which should be at 0.168 cycles/hour) has vanished, casting doubts on
the reality of this modulation. 

More important is, however, the survival
of only one significant peak common to the data sets of both years. A
blown-up view of of the Lomb-Scargle periodogram of a small frequency 
interval around this peak is plotted in the central frames of 
Fig.~\ref{orb-period} and shows the perfect alignment of the signals.
The average of the corresponding periods in the two seasons is 
4.839 h.  
Within the error bars quoted by Nogami et al.\ (1999) it is identical 
to the spectroscopic orbital period, lending credibility to the reality of 
this photometric period which we will henceforth identify with $P_{\rm orb}$. 
The formal error is only
0.00003~h if it is taken as the standard deviation of the two independent
period measurements of 2016 and 2018. This may be an underestimate. If
we take instead the standard deviation of a Gaussian fit to the corresponding
peak in the power spectra the error increases to
0.008~h. Folding the light curves on $P_{\rm orb}$, choosing the epochs such that
the minimum coincides with phase 0, results in graphs plotted in the right
frames of Fig.~\ref{orb-period}. The best fit sine curve is also shown (in
red). Note that it is not a perfect fit, leaving systematic residuals at
maximum and minimum phases. Thus, not unexpectedly, the orbital variations are
more complex than a simple sine wave. At 20.2 mmag in 2016 and17.4 mmag in 
2018 the total amplitude of the sine curves and thus the amplitude of the 
variations are similar in both years.
 
Adopting the more conservative error limit of 0.008~h for $P_{\rm orb}$ 
leads to a cycle count uncertainty of $\sim$1.2 between the two observing 
seasons. Thus, it is not possible to unambiguously connect both data sets.
The period can be expressed as $\Delta T/E$, where $\Delta T$
is the time difference between two minima in the light curve and $E$ is an
integer values, indicating the number of elapsed cycles. Let 
$\Delta T = 747.459$~days 
be the time difference of the epochs chosen to calculate the phase folded
light curves in the right frames of Fig.~\ref{orb-period}. Then only $E=3707$
or $E=3708$ lead to periods which are compatible within its errors with the 
preliminary period quoted above. This implies that either
$P_{\rm orb} = 0.201634\ {\rm d} \hspace{1em} (4.83922\ {\rm h})$ 
or
$P_{\rm orb} = 0.201580\ {\rm d} \hspace{1em} (4.83792\ {\rm h})$.

Using as a conservative criterion for the uncertainty of these 
values the condition that it must not lead to a phase difference of more than
10\% over the time interval $\Delta T$, the period error is 0.0000054~d
(0.00013~h). The minimum epoch is ${\rm BJD_{min}}\ 2458184.975 \pm 0.020$,
where the error was again chosen such that a maximum phase error of
10\% is permitted.

\section{A coherent 26.7 min variation}
\label{A coherent 26.7 min variation}

\begin{figure}
\parbox[]{0.1cm}{\epsfxsize=14cm\epsfbox{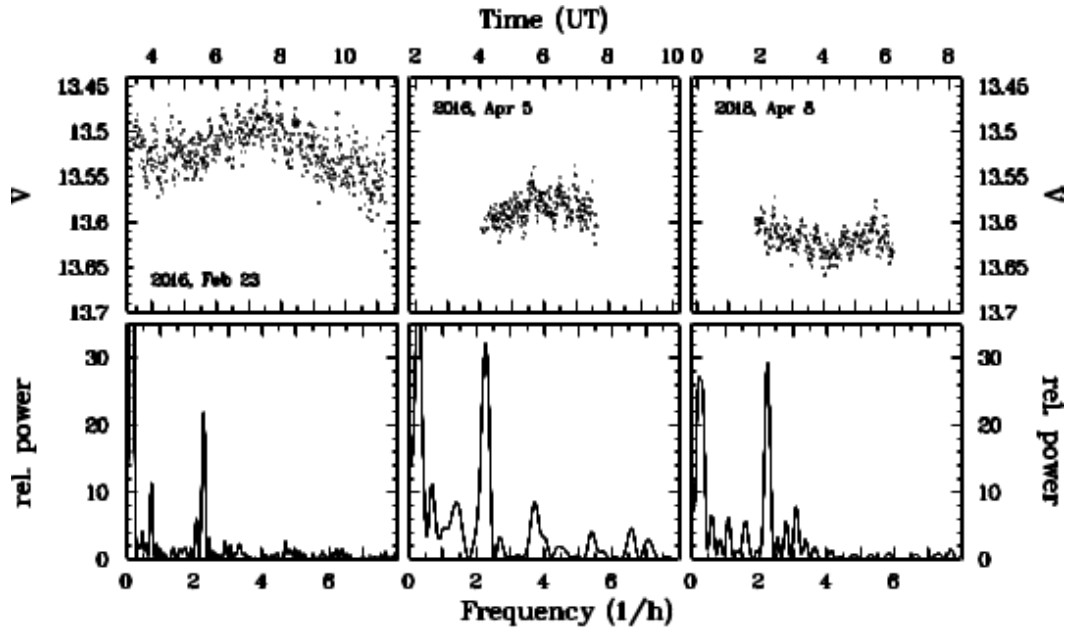}}
      \caption[]{{\it Upper frames:} Three representative light curves
                 of AT~Cnc, all drawn on the same magitude and time scale.
                 {\it Lower frames:} Lomb-Scargle periodograms of the
                 light curves shown in the upper frames, also drawn on
                 the same scale.}
\label{rot-period-1}
\end{figure}

As can be seen in Fig.~\ref{lightcurves}, variations on shorter time
scales are superposed upon the modulations occurring on the time scale
of hours. At first glance they appear to the eye as
irregular flickering, a common feature in all CVs (Bruch 1992).
However, in some cases there seems to be a certain regularity. 
Indeed, regarding the respective frequency range of the Lomb-Scargle
periodograms of the individual nightly light curves we find that in
many of them not only significant signals at frequencies corresponding
to the periodicity of the apparent variations occur, but -- more so -- 
that they always appear at the same frequency of $\sim$2.25 cycles/hour. 
In anticipation of the interpretation of these variations in 
Sect.~\ref{Discussion} we will call the period corresponding to this
frequency the rotational period $P_{\rm rot}$ and thus find 
$P_{\rm rot} = 26.7$~min. The
average frequencies measured during the two observing seasons of 2016 and
2018 are identical within their respective standard deviations. Thus, 
these light changes should not be regarded
as some quasi-periodicity, but as a coherent modulation
of the light of AT~Cnc. In the last but one column of 
Table~\ref{Journal of observations} a plus sign (``+'') indicates those
light curves which definitely exhibit the signal in their power
spectra, while a minus sign (``--'') is assigned to those
lacking it. A ``o'' marks the cases, where a weak
signal is present at the corresponding frequency which, however, would
not be regarded as significant by itself (i.e., without expecting it
to be present). Fig~\ref{rot-period-1} shows three examples of light
curves (upper frames) and their corresponding Lomb-Scargle periodograms
(lower frames). 

In order to test the stability of $P_{\rm rot}$ and to get a more precise
numerical value, those light curves which strongly exhibit the corresponding
signal, were combined into a single data set, separately for the two
observing season. In this case, and differently from the procedure
adopted in Sect.~\ref{Orbital variations}, not only the average magnitude
of the individual light curves was subtracted in order to remove night-to-night
variations and differences of the magnitude scales, but a polynomial of 
suitable order was fit and subtracted from each of them in order to reduce low
frequency noise. The resulting combined light
curves were then again submitted to a frequency analysis.

\begin{figure}
\parbox[]{0.1cm}{\epsfxsize=14cm\epsfbox{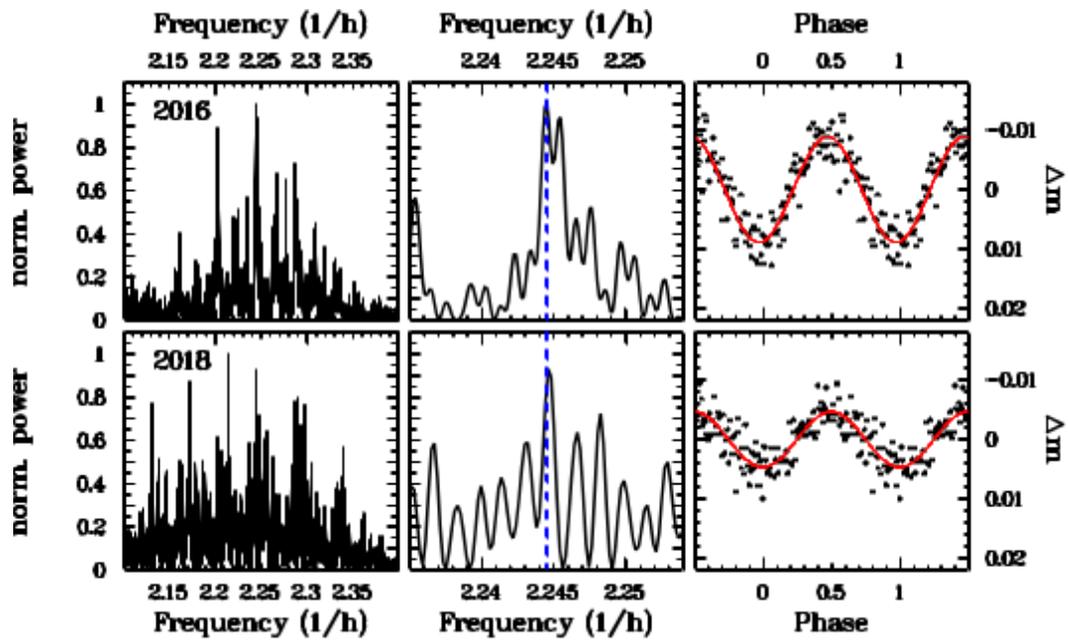}}
      \caption[]{Analysis of the 26.7~min variation in AT~Cnc.
                 {\it Left:} Lomb-Scargle periodograms of the
                 combined light curves of AT~Cnc of 2016 (top) and 2018 
                 (bottom), normalized to the power of the highest peak.
                 {\it Center:} The same periodograms as shown in the left
                 frame, restricted to a narrow range of frequencies around
                 the frequency of the highest peak in the 2016 periodogram.
                 The broken vertical line corresponds to the average of the
                 frequencies of the highest peaks in the both years.
                 {\it Right:} The combined light curves, folded on the
                 period of the highest peak in the central frames, binned
                 in phase bins of width 0.01. The red curves are the best
                 sine fits to the data. (For interpretation
                 of the references to colour in this figure legend, the
                 reader is referred to the web version of this article.)}
\label{rot-period-2}
\end{figure}

The result is convincing and is shown in Fig.~\ref{rot-period-2}. The
left frames show the power spectra for 2016 (top) and 2018 (bottom). 
Again, both contain a complicated pattern of peaks which is to be expected 
considering the uneven time distribution of the individual contributing
light curves. Considering the most significant maxima, there is only one
coincidence between the two observing season: The highest peak in 2016 is
well aligned with the second highest peak in 2018. This becomes even clearer
regarding the central frames of the figure which contain blown-up versions
of the power spectra, restricted to a narrow frequency interval around the
two peaks. The blue vertical lines, which corresponds to the mean of the
frequency maxima of the two peaks, clearly reveals their excellent 
alignment\footnote{The satellite peak to the right of the main peak
in 2016 can be explained as an alias caused by the window function.}. 
We take the standard deviation of a Gaussian fit to the 
peaks as a conservative measure of the frequency error. Then
$P_{\rm rot} = 26.731 \pm 0.005$ min. The combined light curves, folded on
this period and binned in phase intervals of width 0.01 are plotted in
the right frames of Fig.~\ref{rot-period-2}. They can be well represented
by an sine wave, and the red curve in the figure represents the best fit
sine curve which has a full amplitude of 17.8 mmag in 2016 and 9.4 mmag in
2018. Again, the epoch has been chosen such that the minimum
coincides with phase 0, independently for 2016 and 2018.

The error of $P_{\rm rot}$ is too large to permit a reliable cycle count
between the two observing seasons. Over the $\sim$2 years between them
the ambiguity sums up to $\sim$7.5 cycles. Therefore, we can only say
that $P_{\rm rot} = \Delta T/(E + n)$, where $\Delta T$ is the time difference
between the observations, the integer value $E$ is an estimate for the 
elapsed number of cycles, and $n$ is the error of $E$, also expressed as
an integer value. Using 
$\Delta T = 743.083$, i.e., the difference of the epochs used to
calculated the folded light curves in the right frames Fig.~\ref{rot-period-2},
together the preliminary value of $P_{\rm rot}$ as quoted above gives
a best estimate of $E=40029$. Then, $P_{\rm rot} = 26.73161 \pm 0.00007$~min,
where the error was defined such as to permit a maximum phase error of
10\% between the observing reasons. Remember, however, that the true
period may well be an alias of this value!

\section{Discussion}
\label{Discussion}

\subsection{The orbital period}
\label{The orbital period}

This investigation has revealed two consistent periods in the light curves
of AT~Cnc which were observed in two observing seasons separated by two
years. The first of them lies within the error limits of the previously
determined spectroscopic orbital period. We therefore 
identify it as a manifestation of orbital variations. The error of the
photometric period is smaller than that of the spectroscopic one. Thus,
the revolution period of AT~Cnc could be determined with a higher
accuracy, which, however, is unfortunately not high enough to permit 
an unambiguous cycle count between 2016 and 2018 and thus
to increase the accuracy by another order of magnitude. 

The amplitude of the orbital variations is
only of the order of 0.01 mag, significantly smaller than the total range
of variations seen in individual nightly light curves. These are thus
dominated by modulations which must have other origins and which we will
not explore in more detail here. We only mention that this behaviour is
not uncommon in cataclysmic variables. 
Orbital variations in CVs are in general attributed to the changing
visibility of structures in the accretion disk as -- in a frame of 
reference fixed in the binary system -- the line of sight to the observer
changes. The accretion disk being largely a two-dimensional entity aligned
to the orbital plane, this effect should decrease with decreasing orbital
inclination of the system. Nogami et al.\ (1999) have shown that we see
AT~Cnc more or less along the axis of the binary revolution. This is in 
line with the small amplitude of the orbital modulations.  

Apart from $P_{\rm orb}$ no other convincing persistent periodicity on the 
time scale of hours could be seen in our data. Therefore, it can be concluded
that all other periodic variations in the light curves of AT~Cnc as seen
by Goetz (1986), Nogami et al.\ (1999) and Kozhevnikov (2004)
are transient features. In particular, the negative superhumps claimed by
Kozhevnikov (2004), if real, do not persist over time. 

\subsection{The 26.7 min period}
\label{The 26.7 min period}

The second consistent periodicity found here, $P_{\rm rot}$, is a coherent 
oscillation with
a period of about half an hour. It is doubtful if it can be associated to the
broad hump at frequencies of 0.4 -- 0.7 mHz which Kozhevnikov (2004)
saw in his average power spectrum. The period coincides with the upper limit
of this frequency range, but in the present data $P_{\rm rot}$ causes a sharp
peak in the power spectra, not a broad hump. Moreover, the individual power
spectra shown in Fig.~3 of Kozhevnikov (2004) do not contain this 
peak. Two mechanisms come to mind to explain the 26.7~min modulation: white
dwarf oscillations or white dwarf rotation. We will consider these 
possibilities in turn.

\subsubsection{White dwarf oscillations}
\label{White dwarf oscillations}

White dwarf oscillations are not commonly seen in CVs, but they are not
unprecedented. The first and (to our knowledge) so far only confirmed 
case is the WZ~Sge type dwarf
nova GW~Lib. The white dwarf in this system exhibits non-radial oscillations
(van Zyl et al. 2000, 2004) otherwise observed in some isolated white dwarfs 
of tye DA, i.e., the ZZ~Cet stars.

However, an interpretation of the 26.7~min oscillation in AT~Cnc in this way 
has a number of 
difficulties. To begin with, the period is longer than any found in
the comprehensive list of ZZ~Cet star characteristics of Kepler (2016).
Moreover, ZZ~Cet stars often exhibit more than one period. In light curves
obtained in several years, GW~Lib, for example, routinely exhibits two or 
three periods (disregarding the fine structure in the respective power 
spectra; van Zyl et al. 2000) which are not absolutely stable. In contrast,
only a single oscillation period is found in AT~Cnc, at a stable frequency in
all observations obtained so far. 

The oscillation amplitude in ZZ~Cet stars range from a few milli-magnitudes 
to a couple of centi-magnitudes (Kepler 2016). While 
this is compatible with the amplitudes seen in AT~Cnc, those values hold
for naked white dwarfs. But in a CVs the oscillations will be diluted
by the light from the accretion disk. The observed amplitudes in GW~Lib
as observed by van Zyl et al.\ (2000) range between 14.5 mmag and 1.9 mmag,
with values above 10 mmag being rare. This is not unlike what is seen in 
AT~Cnc. However, in a short period WZ~Sge star in
quiescence the accretion disk is expected to be rather faint such that the
white dwarf light contributes significantly to the total visual light.
AT~Cnc at standstill and with a much longer orbital period will have a
much brighter accretion disk. Thus, for the observed oscillation amplitude
to reach almost 20 mmag, the intrinsic amplitude must be much higher\footnote{A
hand waving argument, assuming that the white dwarf and the accretion disk 
have the same brightness in a system with an orbital period of 1.5 h, using 
typical CV component masses and Kepler's law
to scale the system dimensions (and thus the disk surface) to the period 
of AT~Cnc, and considering the higher disk luminosity at standstill, shows 
that the intrinsic oscillation amplitude of the white dwarf must then be
of the order of 0.5 mag, an order of magnitude higher than any 
amplitude observed in a ZZ~Cet star.}.

Finally, the white dwarfs in dwarf novae above the period gap all have
surface temperatures well in excess of 20\,000~K (Sion 2012). This is much
beyond the empirical range of $\sim$11\,000~K to $\sim$12\,000~K
(Van Grootel et al. 2012) of the instability strip within which ZZ~Cet type 
oscillations can occur. 

All these arguments speak against an origin of the 26.7~min period as an
oscillation of the white dwarf in AT~Cnc. 

\subsubsection{White dwarf rotation}
\label{White dwarf rotation}

The second mechanism to explain the 26.7~min oscillation is
rotation of the white dwarf in the framework of an intermediate polar
(IP) model for AT~Cnc. In such
systems the white dwarf has a substantial magnetic field which is able
to disrupt the inner accretion disk and guide matter via a curtain shaped
structure to the magnetic poles, generating bright spots on the compact star. 
The varying aspect of these spots as the white dwarf rotates together with
self occultations by the body of the white dwarf and absorption by (parts of)
the accretion curtain modulates the observed light as the star rotates.

\begin{figure}
\parbox[]{0.1cm}{\epsfxsize=14cm\epsfbox{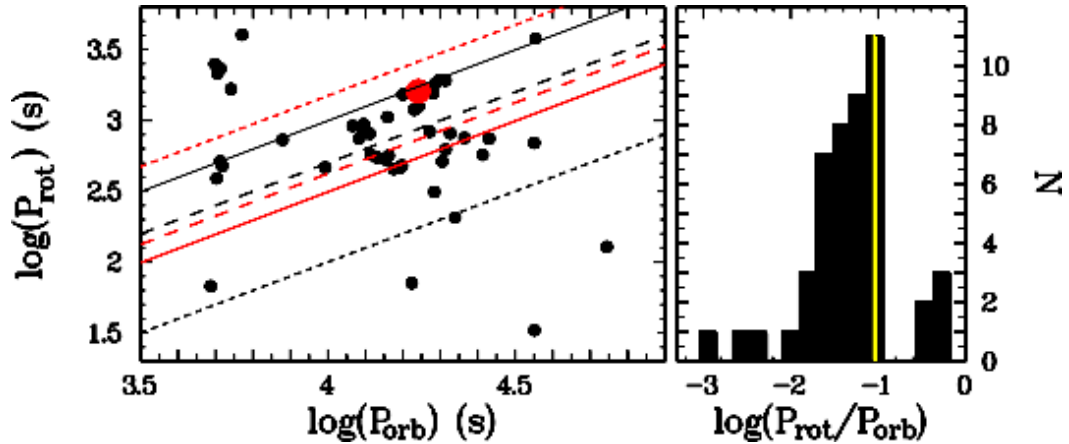}}
      \caption[]{{\it Left:} Location of confirmed intermediate polars in
                 the $P_{\rm orb} - P_{\rm rot}$ plane. The location of AT~Cnc
                 is indicated as a red dot under the assumption that the
                 26.7 min oscillation represents the rotation period of the
                 white dwarf. The solid, dashed and dotted black lines
                 represent $P_{\rm rot}/P_{\rm orb}$ ratios of 0.1, 0.5, and 0.01,
                 respectively. The red lines (solid, dashed and dotted)
                 embody the equilibrium condition according to
                 Eq.~10 of King (1991) for mass ratios of
                 1, 0.5, and 0.1, respectively. 
                 {\it Right:} The distribution of the logarithms of the
                 $P_{\rm rot}/P_{\rm orb}$ ratio of confirmed intermediate
                 polars. The location of AT~Cnc is indicated as a yellow
                 vertical line. (For interpretation
                 of the references to colour in this figure legend, the
                 reader is referred to the web version of this article.)}
\label{pspin-porb}
\end{figure}

In many IPs the ratio of the white dwarf rotation period to the
orbital period is close to 0.1 (Barrett et al. 1988, Warner \& Wickramasinghe
1991). King (1991) show that this is the consequence of an 
equilibrium configuration naturally assumed by IPs if
the white dwarf essentially accretes the specific angular momentum of
the secondary star. In the present case we find $P_{\rm rot}/P_{\rm orb} = 0.092$, 
in excellent agreement. In the left frame of Fig.~\ref{pspin-porb} the
location of all confirmed IPs listed on K.\ Mukai's
IP home page\footnote{https://ads.gfsc.nasa.gov/Koji.Mukai/iphome/iphome.html;
we use all confirmed and ``ironclad'' systems in Mukai's notation.}
in the $P_{\rm orb} - P_{\rm rot}$ plane is plotted (neglecting the long
orbital period system GK~Per). The solid, dashed and dotted black lines
represent $P_{\rm rot}/P_{\rm orb}$ ratios of 0.1, 0.5, and 0.01,
respectively. The red lines (again in the sequence solid, dashed and dotted)
embody the spin equilibrium condition according to Eq.~10 of 
King (1991) for mass ratios of $q = M_2/M_1 = 1$, 0.5, and 0.1, 
respectively. Here, $M_1$ and $M_2$ are the masses of the white dwarf and
the secondary star. The location of AT~Cnc is indicated by a red dot and 
is very well aligned with numerous other intermediate polars. This is
also evident in the right frame of Fig.~\ref{pspin-porb}, which contains
a histogram of the (logarithmic) $P_{\rm rot}/P_{\rm orb}$ ratios. AT~Cnc,
indicated by a yellow vertical line, falls right on the peak of the 
distribution.

The power spectra of some IPs exhibit signals at the orbital side bands
of the white dwarf rotation frequency. 
(e.g., TX~Col, Buckley \& Tuohy 1989; AO~Psc, Hellier et al. 1991).
These may be caused by reprocessing of radiation from the emission regions
on the white dwarf in a region fixed in the orbital frame of reference.
Our periodograms contain signals at 2.4514~h$^{-1}$ (2016) and 
2.4508~h$^{-1}$ (2018). This is close to 
$1/P_{\rm rot} + 1/P_{\rm orb} = 2.4512$~h$^{-1}$. While they may thus be positive
orbital sidebands, they are weak and it would be premature to definitively 
identify them as such.

The amplitude of the $P_{\rm orb}$ variations is small compared to what is 
observed in most IPs (except the small subset of IPs named DQ~Her stars, 
where the white dwarf rotation period is less than $\sim$2 min; see chapter 8
of Warner, 1995). This may be due to the 
low orbital inclination of AT~Cnc: If the magnetic poles are approximately
aligned with the rotation axis, we may see only one pole and the aspect of 
the bright spot on the white dwarf surface may not change much due to
rotation. Consequently, the light modulation and thus the amplitude remains
small.

The amplitude is, however, much smaller in 2018 than is was in 2016. Since
the average system magnitude in both seasons was the same within a few
hundredth of a magnitude, this cannot be explained by a brighter 
accretion disk in 2018 diluting the rotational modulation. Instead, the latter
must have been systematically smaller in 2018. The different strength 
of the respective power spectrum signal in different nights, up to its
complete disappearance also points at significant changes in the flux
of the light source causing the modulations. We may speculate about
a changing mass accretion onto the white dwarf. Either the accretion
decreases to a degree that the bright spot at the visible magnetic pole
gets too faint to cause a sufficient modulation of the total light output
of the system, or -- assuming a weak magnetic field -- an increased mass
transfer causes the Alfv\'en radius, i.e., the radius at which the
accretion is dominated by the magnetic field, to decrease so much that it 
becomes smaller that the white dwarf radius or at least small enough such that
the foot point of the accretion curtain on the white dwarf becomes too large
for significant modulations  due to the changing aspect caused by the 
white dwarf rotation to be seen. However, in both cases this should be
accompanied by a change in the system magnitude: in the first case to
a lower brightness (decrease of accretion luminosity), in the second case
to higher brightness (increase of accretion luminosity). Such a correlation
is not obvious in our data.  

The lack of observed x-rays from AT~Cnc may appear as an obstacle for the
interpretation of the system as an intermediate polar because, in general,
IPs are copious hard x-ray emitters.
In fact, a considerable number of them were first identified through their
periodic x-ray modulations. Verbunt det al.\ (1997) searched the ROSAT XRT-PSPC
All Sky Survey for x-rays from 162 CVs with known or detected binary periods.
Among them was AT~Cnc which was not detected. We are also not aware of any
other x-ray observation of the system. However, the sample of 
Verbunt et al.\ (1997) contains also such well-known intermediate polars 
as FO~Aqr, BG~CMi and DQ~Her which were not detected. Therefore, the 
absence of detected x-rays from AT~Cnc in the ROSAT survey may not be a 
conclusive argument against an IP nature. 

\section{Conclusions}
\label{Conclusions}

The analysis of 37 light curves of the Z~Cam type dwarf nova AT~Cnc obtained 
during two standstills spanning about 7 weeks in 2016 and 6 weeks in 2018, 
comprising
a total of 170 hours of observations, did not confirm the consistent presence
of any of the periods on time scales of hours claimed in previous publications.
In particular, no indications of superhumps as reported by
Kozhevnikov (2004) were detected. In the individual light curves
variations with amplitudes of the order of 0.1 mag are common, but they
are not obviously regular. However, a power spectrum analysis of the combined
light curves of each observing season revealed the presence of a low
amplitude ($\sim$0.02 mag) modulation with the same period in 2016 and 2018
and equal to within the
error margin to the spectroscopic orbital period. It may therefore 
confidently be regarded as a photometric manifestation of revolution of
the binary system. Since the error of this photometric period is smaller 
than of the previously known spectroscopic one, the binary period of AT~Cnc
could be determined with a somewhat higher accuracy.

A second, shorter photometric period of $\sim$26.7 min is evident in the
power spectra of many, albeit not all, light curves in both years. Its 
stability in frequency 
is testified by the presence of the same feature in the combined
seasonal light curves. While the corresponding power spectra are noisy
(in particular in 2016) because of the complicated window spectrum,
making it difficult to identify the exact period or to distinguish between
one coherent period or several quasi-periods caused by, e.g., quasi-periodic
oscillations, one and only one of the significant power spectrum peaks
is present at exactly the same frequency in both, 2016 and 2018. Thus,
this appears to be a coherent modulation in AT~Cnc, stable at least over
the time scale of years. We discuss two scenarios for this modulation.
An oscillation of the white dwarf as observed in the WZ~Sge type star
GW~Lib can probably be discarded. More promising is an explanation in
the context of an intermediate polar model of AT~Cnc, although there may
also be problems with this interpretation in view of the presence and
absence of the 26.7~min variations in different nights without a substantial
change of the system brightness, and the lack of observed x-ray emission.

\section*{Acknowledgements}
We gratefully acknowledge the efforts of those AAVSO observers who 
contributed light curves for this study but are not co-authors, namely 
D.\ Barrett and K.\ Menzies. We also thank the AAVSO as an organization 
for their excellent work to maintain the International Database which is
a treasure trove ready to be exploited.

\section*{References}

\begin{description}
\parskip-0.5ex

\item Barrett, P., O'Donoghue, D., \& Warner, B. 1988, MNRAS, 233, 759
\item Bruch, A. 1992, A\&A, 266, 237
\item Bruch, A. 1993, MIRA --  
      A Reference Guide, Astron.\ Inst.\ Univ.\ M\"unster
\item Buckley, D.A.H., \& Tuohy, I.R. 1989, ApJ, 344, 376
\item Buat-M\'enard, V., Hameury, J.-P., \& Lasota, J.-P. 2001, A\&A, 369, 925
\item Eastman, J., Siverd, R., \& Gaudi, B.S. 2010, PASP, 122, 935
\item G\"otz, W. 1983, IBVS 2363
\item G\"otz, W. 1985, IBVS 2734
\item G\"otz, W. 1986, IBVS 2918
\item G\"otz, W. 1988a, IBVS 3066
\item G\"otz, W. 1988b, IBVS 3208
\item G\"otz, W. 1990, Mitt.\ Ver\"and.\ Sterne, 12, 60
\item G\"otz, W. 1991, Mitt.\ Ver\"and.\ Sterne, 12, 111
\item Hellier, C., Cropper, M., \& Mason, K.O. 1991, MNRAS, 248, 233
\item Horne, J.H., \& Baliunas, S.L. 1986, ApJ, 302, 757
\item Kepler, S.O. 2016, http://astro.if.ufrgs/zzcet.htm
\item King, A.R., \& Lasota, J.-P. 1991, ApJ, 378, 674
\item Kozhevnikov, V.P. 2003, A\&A, 398, 267
\item Kozhevnikov, V.P. 2004, A\&A, 419, 1035
\item Kozhevnikov, V.P. 2007, A\&A, 465, 557
\item Lasota, J.-P. 2001, New Astr.\ Rev., 45, 449
\item Lomb, N.R. 1976, Ap\&SS, 39, 447
\item Meinunger, L. 1981, Mitt.\ Ver\"and.\ Sterne, 9, 59
\item Nogami, D., Masuda, S., Kato, T., \& Hirata, R. 1999, PASJ, 51, 115
\item Ratering, C., Bruch, A., \& Diaz, M.P. 1993, A\&A, 268, 694
\item Roberts, D.H., Leh\'ar, J., \& Dreher, J.W. 1987, AJ, 93, 968
\item Romano, G., \& Perssinotto, M. 1968, Publ.\ Astr.\ Obs.\ Padova, 151, 9
\item Scargle, J.D. 1982, ApJ, 263, 853
\item Shara, M.M., Drissen, L., Martin, T., Alarie, A., \& Stephenson, F.R.
      2017, MNRAS, 465, 739
\item Shara, M.M., Mizusawa, T., Wehinger, P., et al. 2012, ApJ, 758, 121
\item Simonsen, M., Boyd, D., Goff, W., et al. 2014, JAAVSO, 42, 1
\item Sion, E.M. 2012, JASS, 29, 169
\item Smith, R.C., Sarna, M.J., Catal\'an, M.S., \& Jones, D.H.P. 1997,
      MNRAS, 287, 271
\item Thorstensen, J.R., Ringwald, F.A., Wade, R.A., Schmidt, G.D., \&
      Norsworthy, J.E. 1991, AJ, 102, 272
\item Van Grootel, V., Dupret, M.-A., Fontaine, G., et al. 2012, A\&A, 439, A87
\item van Zyl L., Warner, B., O'Donogue, D., et al. 2000, Baltic Astr., 9, 231
\item van Zyl L., Warner, B., O'Donogue, D., et al. 2004, MNRAS, 350, 307 
\item Verbunt, F., Bunk, W.H., Ritter, H., \& Pfeffermann, E. 1997,
      A\&A, 327, 602
\item Warner, B. 1995, Cataclysmic Variable Stars, Cambridge University Press,
      Cambridge
\item Warner, B., \& Wickramasinghe, D.T. 1991, MNRAS, 248, 370      

\end{description}

\end{document}